\documentclass[twocolumn,showpacs]{revtex4}
\usepackage{graphics,epsf,bm}
\def\h4s{$^{\hspace*{0.1em}4}_\Sigma$He }

\def\bk{\mbox{\boldmath $k$}}

\begin{document}
\draft
\title{Equivalent hyperon-nucleon interactions in low-momentum space}
\author{M. Kohno,$^1$ R. Okamoto$^2$, H. Kamada$^2$ and Y. Fujiwara$^3$}
\affiliation{$^1$Physics Division, Kyushu Dental College,
Kitakyushu 803-8580, Japan\\
$^2$Department of Physics, Kyushu Institute of Technology, Kitakyushu 804-8550, Japan\\
$^3$Department of Physics, Kyoto University,\\ Kyoto 606-8502, Japan\\
}

\begin{abstract}
Equivalent interactions in a low-momentum space for the $\Lambda N$, $\Sigma N$
and $\Xi N$ interactions are calculated, using the SU$_6$ quark model potential
as well as the Nijmegen OBEP model as the input bare interaction. Because the
two-body scattering data has not been accumulated sufficiently to determine
the hyperon-nucleon interactions unambiguously, the construction of the potential even
in low-energy regions has to rely on a theoretical model. The equivalent interaction
after removing high-momentum components is still model dependent. Because this
model dependence reflects the character of the underlying potential model,
it is instructive for better understanding of baryon-baryon interactions in the
strangeness sector to study the low-momentum space $YN$ interactions.
\end{abstract}

\maketitle

\section{Introduction}
A low-momentum space nucleon-nucleon ($NN$) interaction has been extensively
studied in recent years \cite{BOG,EP99}. If we start with the full space realistic $NN$
interaction, the discussion is focused on the evaluation of an effective
$NN$ interaction in a low-momentum space which reproduces the same eigenvalues
or $T$-matrices in the low-momentum space as those of the original interaction.
In this paper, we call such an effective interaction in a restricted space as
an {\it equivalent interaction}.
It was demonstrated  \cite{BOG} that after the high momentum components
corresponding to short-range repulsion are renormalized, various $NN$
potential models give essentially the same low-momentum $NN$ interaction.

The construction of an energy-independent equivalent interaction in a model
space has been one of the main subjects in effective interaction theories in various
field. Okubo \cite{OKU} proposed in 1950's a unitary transformation method to eliminate
meson-degrees of freedom to obtain the $NN$ potential. Through 1950's
to 1970's, numerous theoretical works \cite{BH58,FB58,BRAN} have been devoted to
nuclear many-body problems starting from realistic nucleon-nucleon interactions.
In the context of the shell model effective interaction, Suzuki and Lee \cite{SL80,LS80}
formulated a decoupling condition method using a similarity transformation
and an iterative procedure to solve the decoupling equation. The theory behind recent
developments is closely related to this viewpoint.

On the two-body level, the collapse of various realistic $NN$ interactions to
the universal interaction in a low-momentum space $k\leq \Lambda$ with
$\Lambda\simeq 2$ fm$^{-1}$ seems not to be
surprising because of abundant $NN$ scattering data in the corresponding
energy region. In the strangeness non-zero sectors the scattering data is scarce at present,
although studies of the hyperon-nucleon $(YN)$ and hyperon-hyperon $(YY)$
interactions have been accelerated with the experimental progress in hyper-nuclear
physics. Thus the construction of the $YN$ and
$YY$  potentials has to rely on a certain theoretical framework to a
large extent. The Nijmegen group \cite{NIJDF,NIJNS,NSC} has been revising
the $YN$ potentials on the basis
of the one-boson exchange potential (OBEP) picture since the end of 1970's.
The initial hard-core models \cite{NIJDF} were
replaced by the soft-core parameterization \cite{NIJNS,NSC}.
The J{\"u}lich group \cite{HHS,RHS,HM} has also been developing another OBEP
model. In spite of the basic flavor SU$_3$ relations, there are
still much uncertainties in the construction of the $YN$ and $YY$ potentials.

A different unified description was proposed for the octet baryon-baryon
interactions by the Kyoto-Niigata group \cite{FU96a,FU96b}
on the basis of an SU$_6$ quark model, in which the gluonic interaction
in the resonating group method (RGM) for two composite nucleons composed of three
quarks is supplemented by the long-ranged one-boson exchange interaction between quarks.
The most recent model fss2 \cite{FSN} achieves comparable accuracy in the $NN$ sector to
modern realistic $NN$ potentials.
The advantage of the SU$_6$ quark model is that the extension of the potential
parameters determined in the $NN$ sector to the strangeness $S=-1$ and $-2$
sectors seems to be less ambiguous than the OBEP models.
Recently, a novel method based on the chiral effective field theory has been applied to
the strangeness $S=-1$ and $S=-2$ baryon-baryon interactions \cite{CEFT1,CEFT2}. 

Because the present potential models for the $YN$ interaction are not well regulated by
experimental data, the equivalent interaction after removing
the high-momentum components can still be model dependent.
In other words, properties of these equivalent interactions reflect the character of
the underlying potential model.
Thus, it is interesting to compare the low-momentum space $YN$ interactions obtained
from different bare potential models.

The evaluation of low-momentum $\Lambda N$ and $\Sigma N$ interactions
was recently reported by Schaefer {\it et al.} \cite{SWW} for the momentum cut-off
value of $\Lambda\sim 2.5$ fm$^{-1}$, using several Nijmegen NSC potentials \cite{NSC}.
The new feature in the $YN$ interactions is the presence, in most cases, of
the coupling between two or three baryon-channels,
such as $\Lambda N$-$\Sigma N$, $\Xi N$-$\Lambda \Lambda$-$\Sigma\Sigma$,
and $\Xi N$-$\Sigma\Lambda$-$\Sigma\Sigma$ couplings.
In this paper, we present equivalent interactions in the low-momentum
space with $\Lambda=2$ fm$^{-1}$ for the $\Lambda N$, $\Sigma N$ and $\Xi N$
potentials starting from the SU$_6$ quark model bare potentials \cite{FSN}.
We choose this cut-off value as a typical low-momentum scale for which the potential
model dependence is almost eliminated in the case of the $NN$ interaction \cite{BOG}.

The naive definition of the potential based on the RGM formalism leads to an
energy-dependent potential through the norm kernel. Such energy-dependence
is not suited for the equivalent interaction theory. Recently, the method to
eliminate the energy-dependence has been developed \cite{SUZ}, using a
renormalized RGM formalism. We use this prescription for the quark model potential.
There is another complexity inherent
in the RGM formulation of the baryon-baryon interaction, which is the presence
of a Pauli forbidden state in certain channels. The interaction has to be applied
in the space in which the Pauli forbidden state is projected out. In other words,
an orthogonal condition to the forbidden state has to be imposed when solving
the Schr\"{o}dinger equation. This condition is not removed even if the
energy-dependence is eliminated. Fortunately, however, on-shell and half-on-shell
$T$-matrices are not influenced by the redundant component \cite{FKNS}. It is noted
that the Pauli forbidden state appears only in the 1S0 (11)s state in the model space
of the baryon-octet bryon-octet interaction, specified by the Elliott
notation $(\lambda\mu)$.

For the $\Lambda N$ and
$\Sigma N$ cases, we also employ the Nijmegen NSC97f model. Our results with
the NSC97f potential are mostly in accord with those shown
by Schaefer {\it et al.} \cite{SWW}.
After removing high-momentum components, we still observe model dependence,
in particular in $\Sigma N$ channels. The comparison of the results
with the quark model and the OBEP model demonstrates different characters of these
potential models. 

In Sec. II, we recapitulate the basics of the equivalent interaction theory in a model
space. In the framework which was initiated by Suzuki and Lee \cite{SL80,LS80} and developed
afterword by collaborators \cite{SZ82,SO83} with including the generalization \cite{SOEK}
to the case that the unperturbed model-space energies are non-degenerate,
the mapping operator $\omega$ which connects the model space $P$ and the rest $Q$ space
plays an essential role to explicitly calculate the equivalent interaction. The mapping
operator $\omega$ is determined by the decoupling condition, which leads to a non-linear
equation for $\omega$. This operator $\omega$ can also be obtained by a linear equation
by using the knowledge of the half-on-shell $T$-matrices, as derived by Epelbaum {\it et al.}
\cite{EP99}. We show, in Sec. II,
a compact derivation of the linear equation. The actual
evaluation of the equivalent interaction can be carried out by discretizing the
entire momentum space as  was argued in the paper by Fujii {\it et al.} \cite{FUJI}.
We extend this method to the case of $YN$ interactions, in which there are
couplings among several baryon-channels.

Results of numerical calculations are presented in Sec. III. We show only
the diagonal matrix elements. Because the single-particle (s.p.) potential of the hyperon
in infinite nuclear matter is determined by them in the lowest order,
they can provide useful information about properties of the hyperon-nucleon interaction.
We first give the results for the $NN$ interaction for comparison, then
present $\Lambda N$ and $\Sigma N$ equivalent interactions in the
low-momentum space with $\Lambda =2$ fm$^{-1}$ calculated both for fss2
and NSC97f. Equivalent interactions for $\Xi N$ are evaluated only for fss2.
Section IV summarizes the results of the present paper.

\section{Equivalent interaction}
We recapitulate basic elements of the equivalent interaction theory, following
the method by Suzuki and Lee \cite{SL80,LS80}. Let us denote the original Hamiltonian
in the entire space by $H=H_0+V$, and divide the full Hilbert space into a model space
and the rest which are denoted by their relevant projection operators $P$ and $Q$,
respectively; $P+Q=1$. In particular, the low-momentum space in the two-body
problem is defined as
$P= \int_{|\bk|<\Lambda} d\bk\: |\bk \rangle \langle \bk |$. The problem is to find the
equivalent operator $H_{eff}$ in the model space $P$, which reproduces the
eigenvalue as those of the original $H$. A formal approach has been well known
as the Feshbach projection method \cite{FB58}.
It is instructive to present the Schr\"{o}dinger equation
$H |\Psi\rangle = E|\Psi\rangle$ in a matrix form.
\begin{equation}
 \left( \begin{array}{cc}
            PHP, PHQ \\
            QHP, QHQ
        \end{array} \right)
 \left( \begin{array}{c}
            P|\Psi\rangle \\
            Q|\Psi\rangle
        \end{array} \right)=E
 \left(
\begin{array}{c}
            P|\Psi\rangle \\
            Q|\Psi\rangle
  \end{array} \right).
\end{equation}
It is straightforward to obtain the following equation in the model space $P$
by eliminating $Q\Psi$ in terms of $Q\Psi=\frac{1}{E-QHQ} QHP\Psi$:
\begin{equation}
P\left\{H+HQ\frac{1}{E-QHQ} QH\right\}P\Psi=EP\Psi .
\end{equation}
Thus $H_{eff}$ can be identified with
\begin{equation} H_{eff}=PH_0P + PVP+ PVQ\frac{1}{E-QHQ} QVP. \end{equation}
In this case we obtain the energy-dependent effective interaction in the model space.
This expression is rather formal, since it includes the energy $E$ to be solved.
It is pedagogical to remark that if $V$ has the property of $QVP=0$ or
$PVQ=0$, $PH_0P+ PVP$ is nothing but $H_{eff}$ from the beginning.

Suzuki and Lee \cite{SL80,LS80} proposed in 1980 the way to construct the
energy-independent equivalent potential in the context of the similarity transformation.
Their consideration serves as the basic
for the recent development of various effective interaction theories. It is elementary to
observe that the eigenvalues of the original Hamiltonian $H$ do not change
when $H$ is transformed by a similarity transformation, namely by a regular matrix $X$ and
its inverse $X^{-1}$ as $H\Rightarrow H' \equiv X^{-1}HX$.
As noted above, if $QX^{-1}HXP=0$ holds, $PX^{-1}HXP$ becomes the equivalent
interaction in the model space $P$. Thus the task to find $H_{eff}$ is reduced to
determine $X$ which satisfies $QX^{-1}HXP=0$.

It is sufficient first to consider a regular matrix $X$ in the following form.
\begin{equation} X=\left( \begin{array}{cc}
            1, 0 \\
            \omega, 1
        \end{array} \right), \hspace{1em}\mbox{then}\hspace{1em}
        X^{-1}=\left( \begin{array}{cc}
            1, 0 \\
            -\omega, 1
        \end{array} \right).
\end{equation}
The mapping matrix $\omega = Q\omega P$, which connects the $P$ and $Q$ spaces, plays a
central role in the following.
The decoupling condition $QX^{-1}HXP=0$ now reads:
\begin{equation} QVP+QHQ\omega -\omega PHP-\omega PVQ\omega =0. \end{equation}
Because this is a non-linear equation for $\omega$, we have to use some iteration method
to solve it. Determining the mapping operator $\omega$,
we obtain an energy-independent equivalent interaction in the model space $P$ as
$PX^{-1}HXP= PH_0P +PV(1+\omega)P$. This equivalent Hamiltonian is not hermitian
at this stage. If we utilize a unitary matrix $\tilde{X}$ in the following Okubo
form \cite{OKU,SZ82} constructed from $\omega$ of Eq. (5) to transform the original $H$,
\begin{equation}\tilde{X} = \left( \begin{array}{cc}
            1, -\omega^\dagger \\
            \omega, \hspace{0.5em}1
        \end{array} \right) \left(\begin{array}{cc}
            1+\omega^\dagger \omega ,0\hspace*{1.5em}\\
            \hspace*{1.5em}0, 1+\omega \omega^\dagger
        \end{array}  \right)^{-1/2},
\end{equation}
the equivalent Hamiltonian is apparently hermitian. The general
argument of constructing an hermitian equivalent interaction was given in ref. \cite{OFS}.
It has to be stressed that other operators corresponding to physical
observables are also transformed accordingly.

Although the equation (5) for $\omega$ is non-linear, Epelbaum {\it et al.} showed
in ref. \cite{EP99} that the linear equation can be set up for the mapping operator $\omega$
by the use of the half-on-shell $T$-matrices in the case of the two-body problem.
This linear equation for $\omega$ is easily derived in the above context
as follows. By definition, the equivalent interaction in the $P$ space
$H_0 +PV(1+\omega) P$ is to reproduce the original $T$-matrices in the $P$ space.
Namely we expect that the following equation should hold.
\begin{equation}
 PTP=PV(1+\omega )P + PV(1+\omega )\frac{P}{\epsilon -H_0}TP,
\end{equation}
where $\epsilon$ is the on-shell energy and we are considering half-on-shell
$T$-matrices. The original equation for $T$ in the $P$ space is
\begin{eqnarray}
 & &PTP= PVP +PV \frac{1}{\epsilon -H_0}TP \nonumber \\
       &=& PVP +PV \frac{P}{\epsilon -H_0}TP+PV \frac{Q}{\epsilon -H_0}TP 
\end{eqnarray}
Comparing Eq. (8) with Eq. (9), we obtain
\begin{equation}
  PV\omega P + PV\omega \frac{P}{\epsilon -H_0}TP=PV \frac{Q}{\epsilon -H_0}TP
\end{equation}
Taking out the common $PVQ$, we are led to the linear equation for $\omega$:
\begin{equation}
  Q\omega P = \frac{Q}{\epsilon -H_0}TP -Q\omega \frac{P}{\epsilon -H_0}TP.
\end{equation}

\section{Results}
When the role of the mapping operator $\omega$ is figured out, we can devise
a direct calculational procedure to determine $\omega$, as was explained as
method-2 and used in Ref. \cite{FUJI}. The extension to the case in
which several baryon-channels couple each other, e.g., the treatment of the
$\Lambda N$-$\Sigma N$ transition, is straightforward. In principle, there is no
difference in the calculational method for the tensor coupling and the
baryon-channel coupling. We only have to enlarge the dimension of the
relevant Hilbert space.
It is noted that in the hyperon-nucleon interaction such a coupling between
$^1P_1$ and $^3P_1$ channels also appears through the antisymmetric
spin-orbit interaction which is absent in the $NN$ case, although we do
not examine the $P$-waves in the following.

We calculate equivalent $\Lambda N$, $\Sigma N$ and $\Xi N$ matrix elements in the
low-momentum space with the cut-off value of $\Lambda=2.0$ fm$^{-1}$
for the $^1S_0$ and $^3S_1$ partial waves, starting from the Kyoto-Niigata SU$_6$
quark model potential fss2 \cite{FSN}. This momentum scale should be
regarded as a representative one for which the potential model
dependence of the description of high momentum components has
been shown \cite{BOG} to disappear in the case of the $NN$ interaction.
The Nijmegen NSC97f \cite{NSC} is also
used for the $\Lambda N$ and $\Sigma N$ interactions.
These two potentials are especially
different in the manner of constructing the short-range part. The Nijmegen
OBEP model is based on the heavy meson exchange picture, while the quark
model uses a RGM framework for nonrelativistic quark-clusters.
As noted in Introduction, the energy-dependence from the RGM kernel is
renormalized to give an energy-independent potential \cite{SUZ}.

Although the calculational procedure is straightforward to treat the baryon-channel
coupling, we have to be careful to interpret the matrix elements of the equivalent
interaction in the $P$ space. To obtain physically meaningful quantities such as
hyperon s.p. energies in nuclear medium, we have to further solve the
baryon-channel coupling problem in the $P$ space. If the coupling effect in the
$P$ space is important, the matrix elements before solving the coupling problem
give no clear physical insight. We encounter such
a typical example in the $\Xi N$ $T=1$ channel.

\subsection{$NN$ interaction}

\begin{figure}
\epsfxsize=8cm
\epsfbox{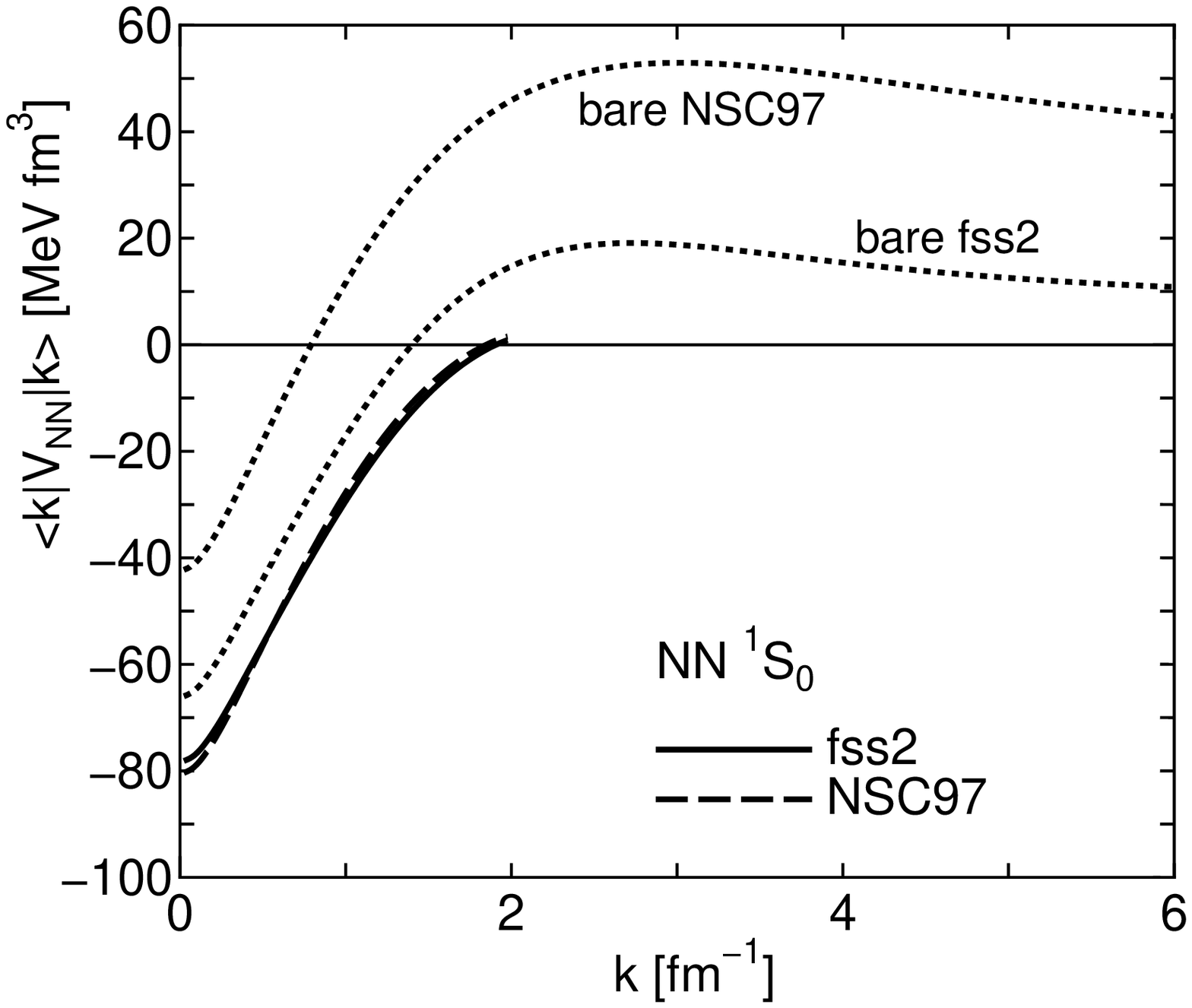}
\caption{Diagonal matrix elements of the equivalent interaction in the low-momentum
space with $\Lambda=2$ fm$^{-1}$ for the $NN$ $^1S_0$ partial wave, using the quark model
potential fss2 \cite{FSN} and the Nijmegen potential NSC97 \cite{NSC}.
Bare matrix elements are shown by dotted curves. 
}
\bigskip

\epsfxsize=8cm
\epsfbox{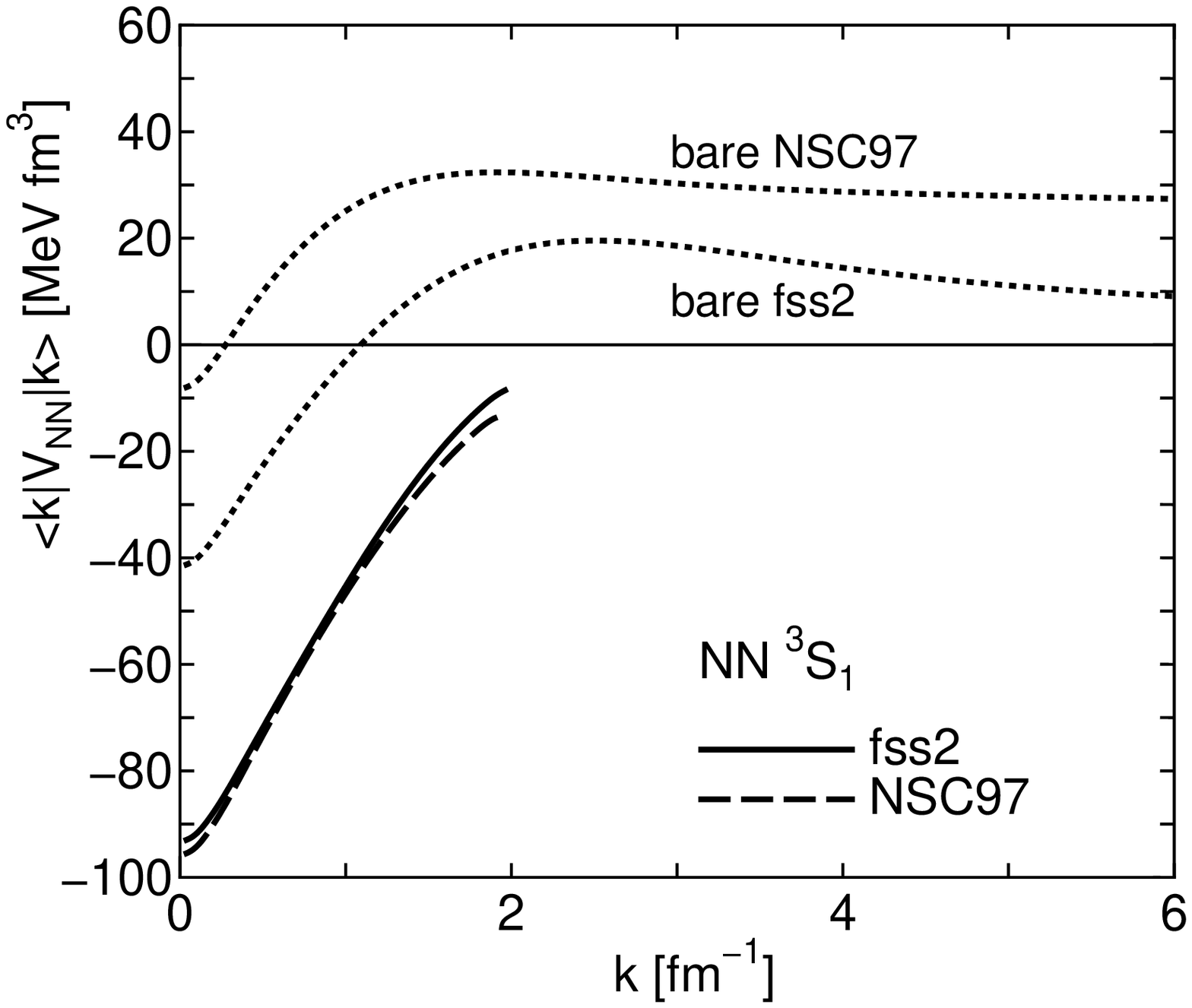}
\caption{Same as Fig. 1, but for the $NN$ $^3S_1$ partial wave. 
}
\end{figure}

Before showing the equivalent interactions in the low-momentum space for the
hyperon-nucleon sectors, we present results of the $NN$ interaction for comparison.
Diagonal matrix elements of the quark model potential fss2 \cite{FSN} in the low-momentum
space with $\Lambda=2$ fm$^{-1}$ are shown by solid curves in Figs. 1 and 2
for $^1S_0$ and $^3S_1$, respectively. These results are very close to those of the
Nijmegen potential NSC97f depicted by dashed curves.
It demonstrates that the quark model potential fss2 achieves the same level accuracy
as modern realistic $NN$ interactions, although the description at the short-range part
is considerably different due to the RGM treatment. The thin curves in Figs. 1 and 2
for the bare matrix elements illustrate that the short-range repulsion of fss2 is rather moderate.

\subsection{$\Lambda N$ interaction}
\begin{figure}
\epsfxsize=8cm
\epsfbox{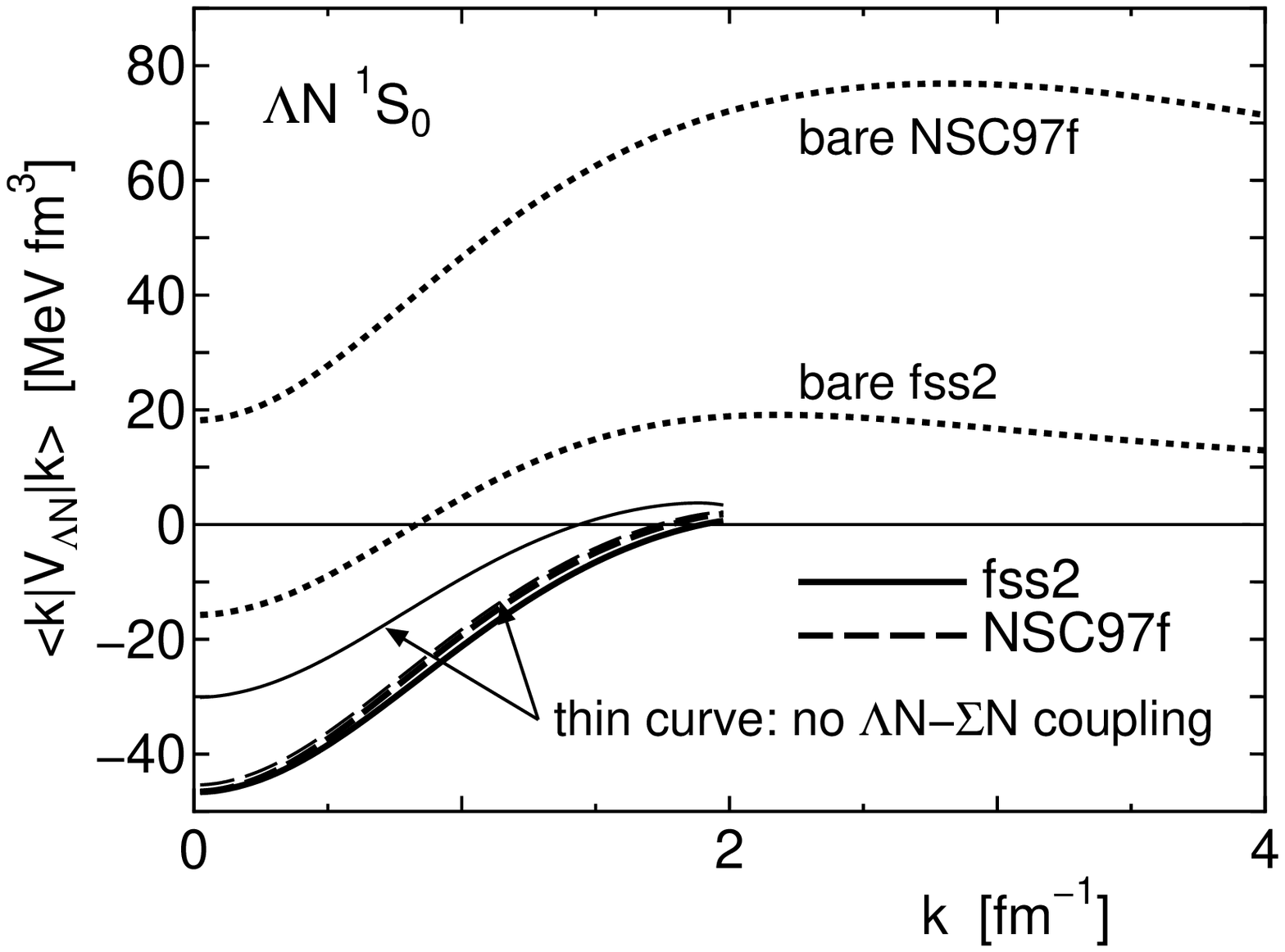}
\caption{Same as Fig. 1, but for the $\Lambda N$ $^1S_0$ partial wave, using fss2 \cite{FSN} and
NSC97f \cite{NSC}. Thin curves are results without $\Lambda N$-$\Sigma N$ coupling.
Thick and thin curves for NSC97f are hard to be distinguished in the figure.
}
\bigskip
\epsfxsize=8cm
\epsfbox{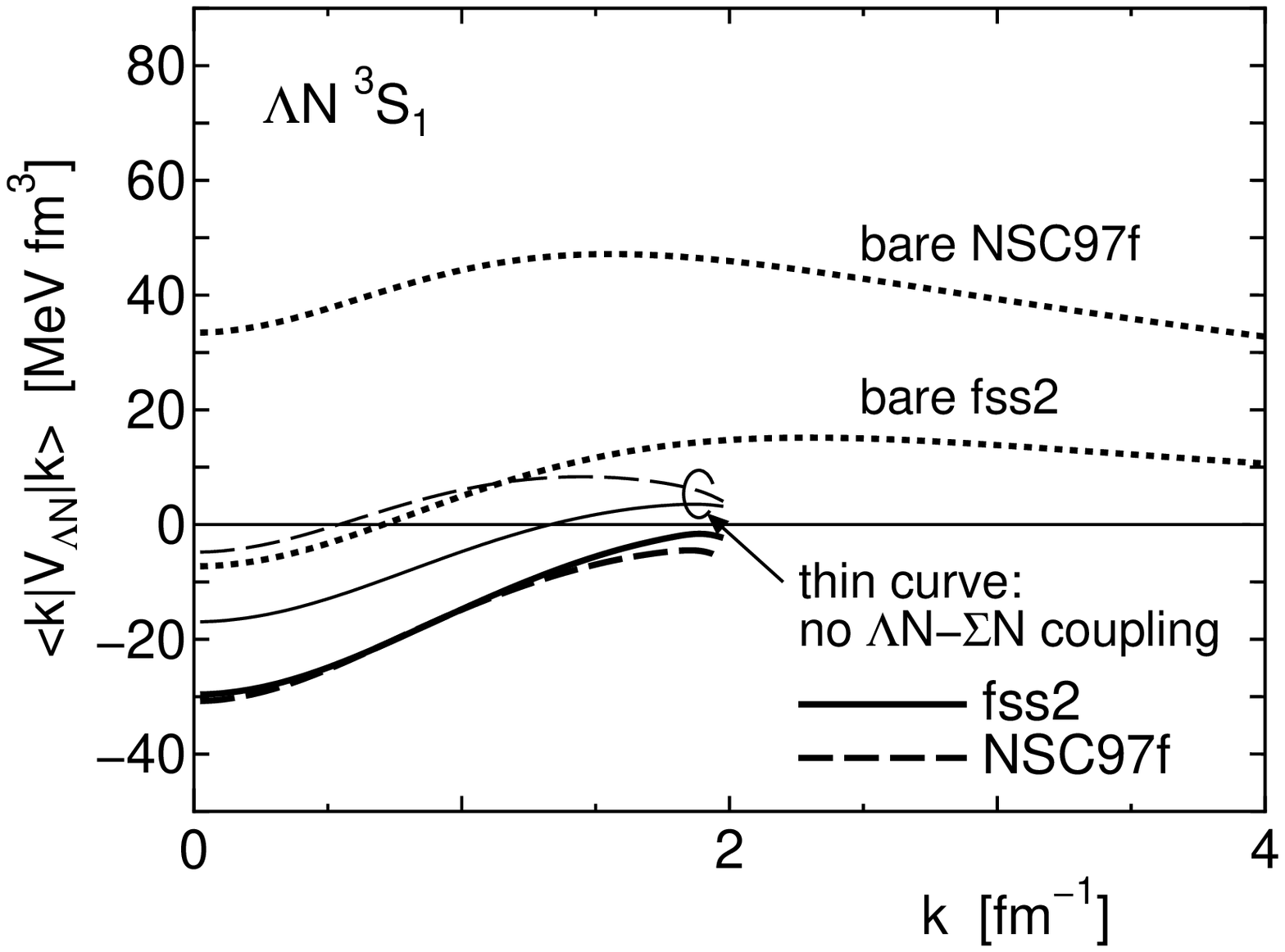}
\caption{Same as Fig. 3, but for the $\Lambda N$ $^3S_1$ partial wave.
}
\end{figure}

Properties of the $\Lambda N$ interaction has been known empirically to some extent,
based on the scattering data from bubble chamber experiments and the data of $\Lambda$
hypernuclei in last 30 years. The central part is fairly well known from the universal depth of the
$\Lambda$-nucleus potential of about 30 MeV, from light to heavy $\Lambda$-hypernuclei,
while the non-central component is still ambiguous. Although the smallness of the $\Lambda$
s.p. spin-orbit splitting in hypernuclei has been established experimentally \cite{TAM},
its relation to the two-body $\Lambda N$ interaction has not been settled. One of the possible
explanations has been offered by the quark model \cite{OSY,FSN} by pointing out the role of
the antisymmetric spin-orbit interaction which can almost cancel the contribution from the
ordinary spin-orbit interaction. In this paper, however, we do not discuss this subject, because
we present the results only for $S$-waves.

Figures 3 and 4 show the low-momentum space diagonal matrix elements of
the equivalent $\Lambda N$ interaction in the $^1S_0$ and $^3S_1$ channels
with $\Lambda=2$ fm$^{-1}$ together with bare matrix elements,
both for the quark model potential fss2 \cite{FSN} and the Nijmegen potential NSC97f \cite{NSC}. 
In order to see the effects of the $\Lambda N$-$\Sigma N$ coupling, we also present
results with turning off the coupling potential by the thin curves.

The Nijmegen potential and the quark model potential provide
very similar matrix elements in the low-momentum space, though the bare matrix elements
are different from each other, reflecting the different character in the
short-range part. Again the short-range repulsion of fss2 is moderate.

In the $^1S_0$ channel, the effect of the $\Lambda N$-$\Sigma N$ coupling is seen to be
negligibly small for NSC97f. Because the pion-exchange is absent in this partial wave,
the $\Lambda N$-$\Sigma N$ coupling is expected to be weak. In the quark model description,
the situation is somewhat different. In the RGM treatment of the two quark-clusters the
Pauli forbidden state appears in this $^1S_0$ $\Lambda N$-$\Sigma N$ channel.
When the $\Lambda N$-$\Sigma N$ coupling is switched off, the orthogonality to the
Pauli forbidden state on the quark level is not strictly satisfied, which may cause an artificial
coupling effect observed in Fig. 3. This feature deserves future investigation. The $^3S_1$
channel is free from the Pauli
forbidden state. Comparing the thick and thin curves in Fig. 4, we see that the considerable
amount of the attractive contribution comes from the $\Lambda N$-$\Sigma N$ coupling
in the outer space beyond $\Lambda=2$ fm$^{-1}$.

It is remarkable to see that the two potential models give almost identical diagonal matrix
elements in the low-momentum space both for $^1S_0$ and $^3S_1$ channels,
though the bare matrix elements are different.
This implies that the potential model for the $\Lambda N$ $S$-wave
interaction is now almost under control. The study of $P$-waves, in particular the
magnitude of the spin-orbit and antisymmetric spin-orbit components, is an important
future subject. The careful study of $\Lambda N$-$\Sigma N$ coupling effects, including
those within the low-momentum space, is also required, because
hypernuclear $\gamma$-spectroscopic measurements \cite{TAM} with
the accuracy of the order of keV are beginning to provide the data of
excitation spectrum of $\Lambda$ hypernuclei.

\subsection{$\Sigma N$ interaction}
\begin{figure}[t]
\epsfxsize=8cm
\epsfbox{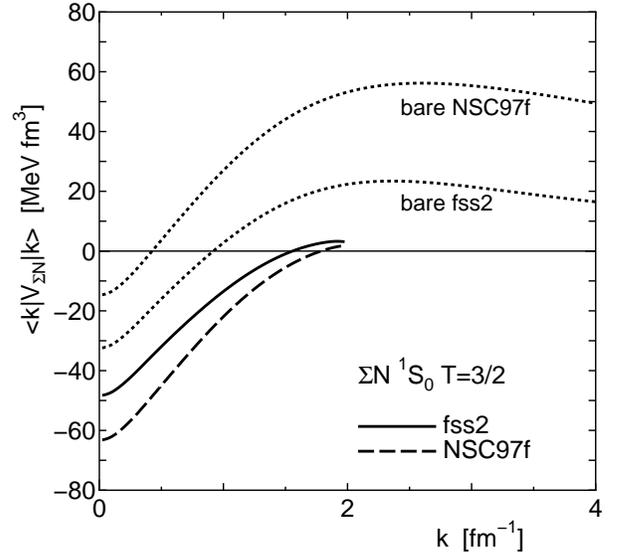}
\caption{Same as Fig. 1, but for the $\Sigma N$ $^1S_0$ partial wave, using fss2 \cite{FSN}
and NSC97f \cite{NSC}.  In this partial wave, there is no baryon-channel coupling.
}
\end{figure}

\begin{figure}[t]
\epsfxsize=8cm
\epsfbox{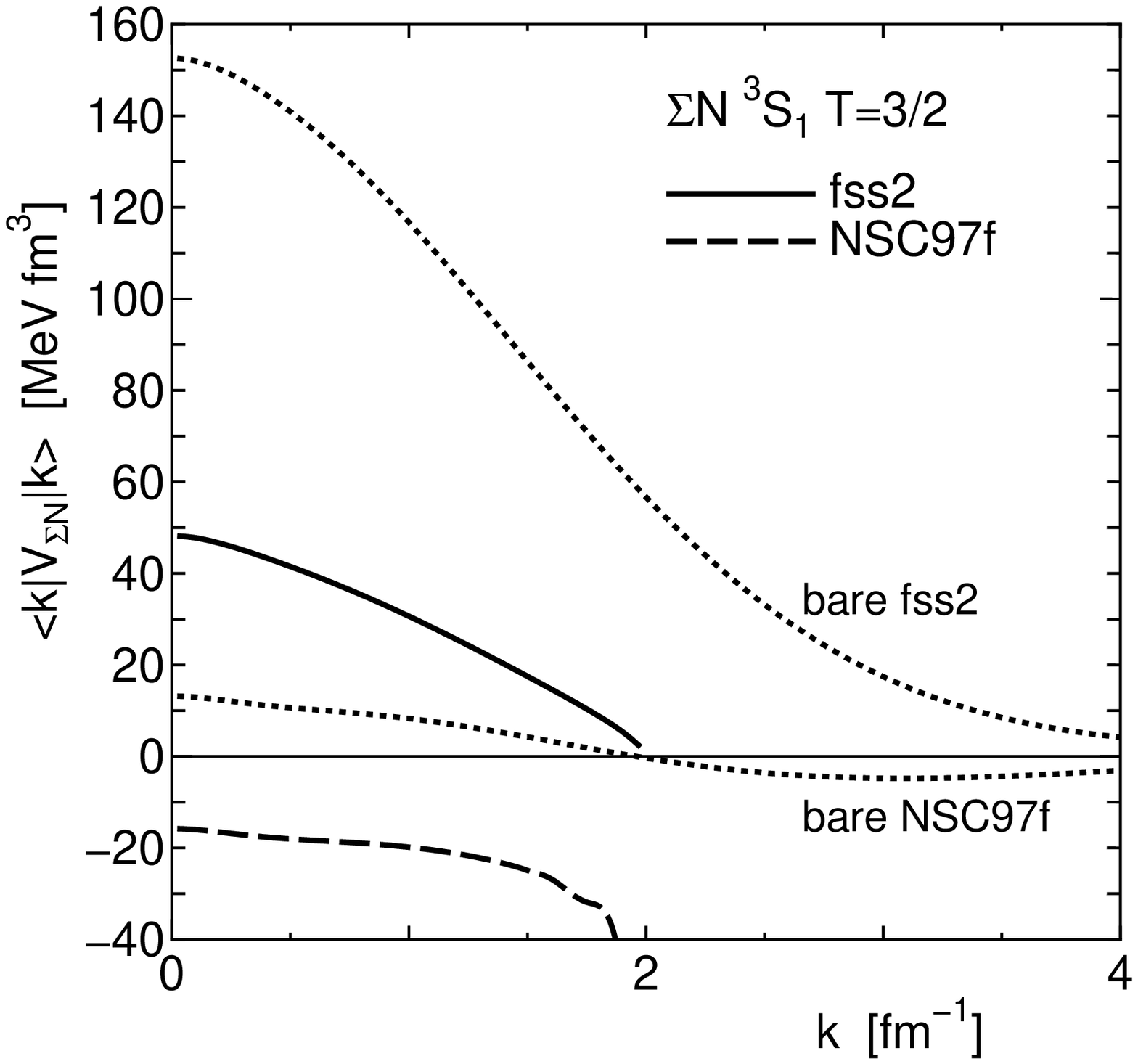}
\caption{Same as Fig. 5, but for the $\Sigma N$ $^3S_1$ $T=3/2$ partial wave.
In this partial wave, there is no baryon-channel coupling.
}
\end{figure}

\begin{figure}[t]
\epsfxsize=8cm
\epsfbox{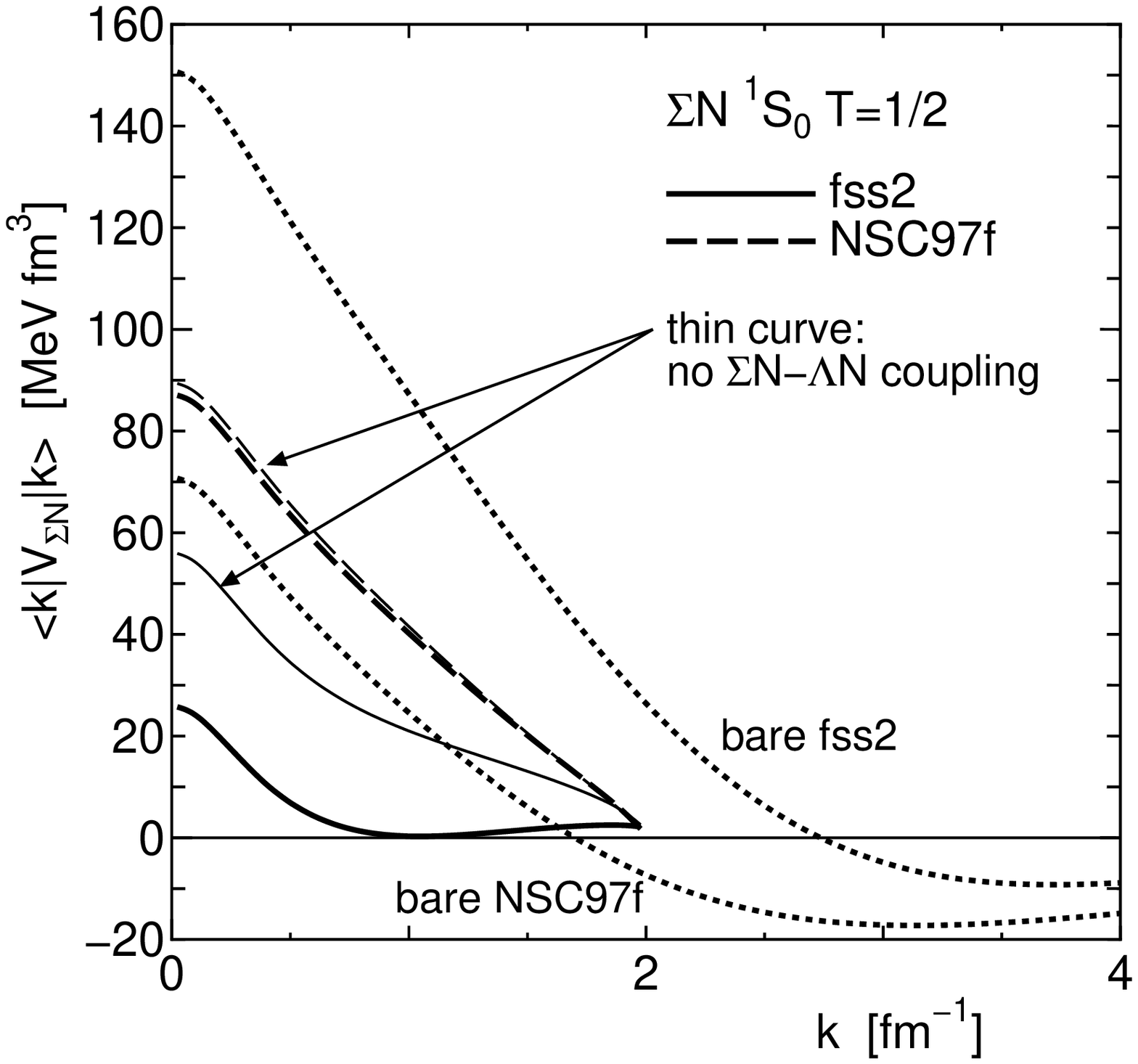}
\caption{Same as Fig. 5, but for the $\Sigma N$ $^1S_0$ $T=1/2$ partial wave.
Thin curves are results without the $\Lambda N$-$\Sigma N$ coupling.
}
\end{figure}

\begin{figure}[ht]
\epsfxsize=8cm
\epsfbox{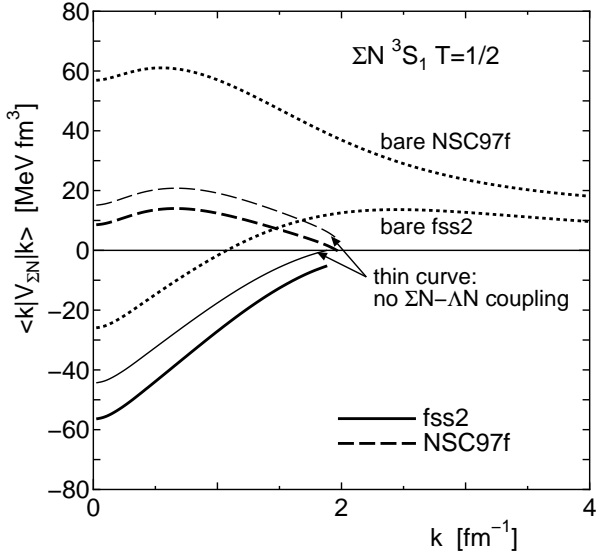}
\caption{Same as Fig. 5, but for the $\Sigma N$ $^3S_1$ $T=1/2$ partial wave.
Thin curves are results without the $\Lambda N$-$\Sigma N$ coupling.
}
\end{figure}

The $\Sigma N$ interaction has been expected to have specific spin and isospin
dependences. Harada {\it et al.} \cite{HAR} demonstrated that the net attractive interaction
in the $^1S_0$ $T=3/2$ channel can yield the bound $^4_\Sigma$He
hypernucleus with $J^\pi =0^+$, which corresponds to the experimental
observation in the $^4$He$(K^-,\pi^-)$ reactions with the stopped kaon \cite{HAY}
as well as the kaon in flight \cite{NAG}. However, probably due to the repulsion in the $^3S_1$
$T=3/2$ channel, $\Sigma$ bound states are unlikely in heavier nuclei, which
has been supported by experimental results on targets $^6$Li and $^9$Be \cite{BNL}.

The quark model picture has been known from the earlier studies \cite{OSY} to give
a definite prediction that certain partial waves such as the $\Sigma N$ $^3S_1$ $T=3/2$
state should be strongly repulsive due to the quark Pauli effect, which has no
explicit counterpart in the OBEP model. The character related to the Pauli effect
might be uncovered in the low-momentum space equivalent interaction.

The $T=3/2$ channels have no $\Sigma N$-$\Lambda N$ coupling because the
$\Lambda$ hyperon is isosinglet. Figure 5 shows the diagonal matrix elements of
the low-momentum space equivalent $\Sigma N$ interaction in the $^1S_0$ $T=3/2$
channel. Results for fss2 \cite{FSN} and NSC97f \cite{NSC} are shown by the solid and
dashed curves, respectively, while bare matrix elements up to $k=4$ fm$^{-1}$ are
shown by the dotted curves. 
The tendency of how high-momentum components are renormalized
is similar to the $NN$ $^1S_0$ case. The attractive interaction in this channel is
important for the existence of the bound $^4_\Sigma$He state. The quark
model potential is sufficiently attractive to yield the bound state as the Nijmegen
potential does. In the viewpoint of the quark model \cite{FSN}, the flavor
SU$_3$ symmetry tells that this channel is dictated by the (22) symmetric
component of the Elliott notation $(\lambda\mu)$. This component is the
same as in the $^1S_0$ ($T=1$) $NN$ state, and thus the rather strong
attractive character after renormalizing high-momentum components is reasonable.

Quantitatively different results are obtained between fss2 and
NSC97f in the $^3S_1$ $T=3/2$ channel, which are shown in Fig. 6.
The quark model potential predicts repulsive interaction, which comes from the
quark Pauli effect. The correlation due to the Pauli effect is not so short-ranged,
as the magnitude of the matrix element at $k=4$ fm$^{-1}$ indicates. The repulsive
character persists in the low-momentum space. On the other hand, the NSC97f model
does not have strong repulsion and the low-momentum equivalent interaction is
attractive. Owing to the spin and isospin weight factors, this $^3S_1$ $T=3/2$ state
dominantly contributes to the $\Sigma$ s.p. potential in nuclear
medium. Analyses \cite{NOUM,HH,MK} of the $(\pi^-,K^+)$ $\Sigma$ formation inclusive
spectra \cite{NOUM} have indicated that the $\Sigma$-nucleus mean field is repulsive.

Figure 7 shows results for the $^1S_0$ $T=1/2$ state. The thin
solid curve presents the result for which the $\Sigma N$-$\Lambda N$
channel-coupling is neglected. The corresponding result for NSC97f
is shown by the dashed curve. Bare matrix elements are also shown up to $k=4$
fm$^{-1}$. Both potential models predict repulsive interaction in this channel.
As we have already seen in the $\Lambda N$ interaction, the $\Sigma N$-$\Lambda N$
channel-coupling effect is very weak in NSC97f. It is noteworthy that the diagonal
matrix elements in the low-momentum space become more repulsive in NSC97f
when renormalizing the high-momentum attractive components.

Figure 8 shows results for the $^3S_1$ $T=1/2$ state. Here we see characteristic
difference for the prediction of the $\Sigma N$-$\Lambda N$ coupling effect
between fss2 and NSC97f. The behavior of the
diagonal matrix elements with fss2 resembles that of the $NN$ $^3S_1$ state
given in Fig.2. In fact, from the flavor SU$_3$ symmetry this channel has a (03)
component by 50\%, which is the component of the $NN$ $^3S_1$ interaction.
Because the other half component of the (11)$_a$ gives a minor contribution,
the $\Sigma N$ interaction in the $^3S_1$ $T=1/2$ state should be similar to
the $NN$ $^3S_1$ interaction in the quark model description.
In contrast, weakly repulsive matrix elements are seen for the $^3S_1$ $T=1/2$
$\Sigma N$-$\Sigma N$ interaction from NSC97f. It has to be remarked, however, that
the $\Sigma N$-$\Lambda N$ coupling in the $P$ space,
which is not yet taken into account, can change even the sign of the diagonal
matrix elements. In fact, $G$-matrix calculations \cite{KF} in symmetric nuclear matter
with using NSC97f tells that the contribution from this channel to the $\Sigma$
s.p. potential in nuclear medium is attractive and becomes repulsive if we switch off the
$\Sigma N$-$\Lambda N$ coupling. This fact simply means that we have to be
careful to interpret the matrix elements in the case that there exist coupling
channels and the coupling effect is important in the low-momentum space.
We encounter such a situation also in the $\Xi N$ $^1S_0$ $T=1$ state.

\subsection{$\Xi N$ interaction}
\begin{figure}[t]
\epsfxsize=8cm
\epsfbox{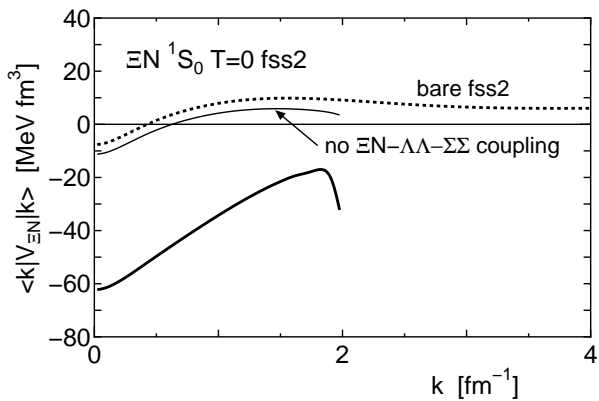}
\caption{Diagonal matrix elements of the equivalent interaction in the low-momentum
space with $\Lambda=2$ fm$^{-1}$ for the $\Xi N$ $^1S_0$ $T=0$ partial wave,
using fss2 \cite{FSN}. Bare matrix elements are shown by a dotted
curve. The thick and thin curves are results with and without taking into account the
$\Xi N$-$\Lambda\Lambda$-$\Sigma\Sigma$ channel coupling, respectively.
}
\end{figure}

\begin{figure}[ht]
\epsfxsize=8cm
\epsfbox{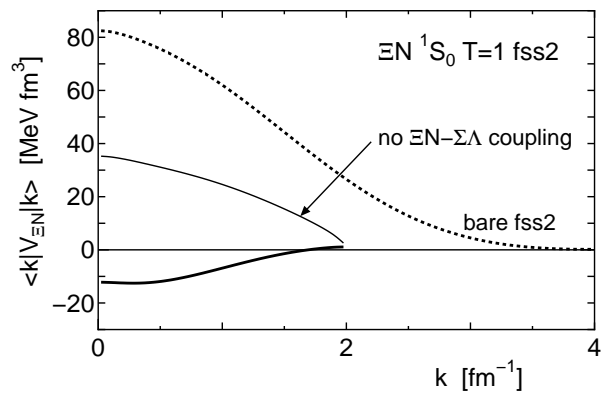}
\vspace*{-3mm}
\begin{center}(a)\end{center}
\vspace*{3mm}
\epsfxsize=8cm
\epsfbox{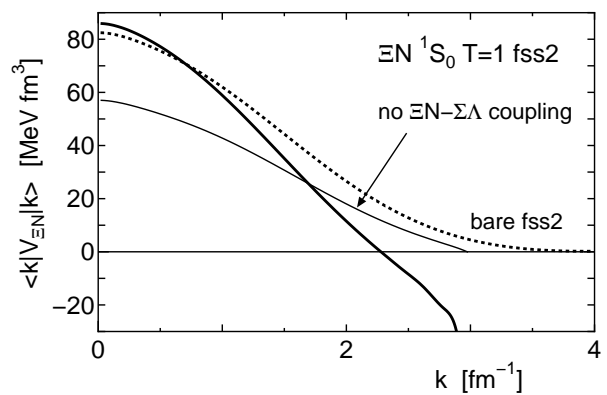}
\vspace*{-3mm}
\begin{center}(b)\end{center}
\vspace*{-2mm}
\caption{Same as Fig. 9, but for the $\Xi N$ $^1S_0$ $T=1$ partial wave.
(a) with $\Lambda=2$ fm$^{-1}$ and (b) with $\Lambda=3$ fm$^{-1}$.
The thick and thin curves are results with and without taking into account the
$\Xi N$-$\Sigma\Lambda$ channel coupling, respectively.
}
\end{figure}

\begin{figure}[ht]
\epsfxsize=8cm
\epsfbox{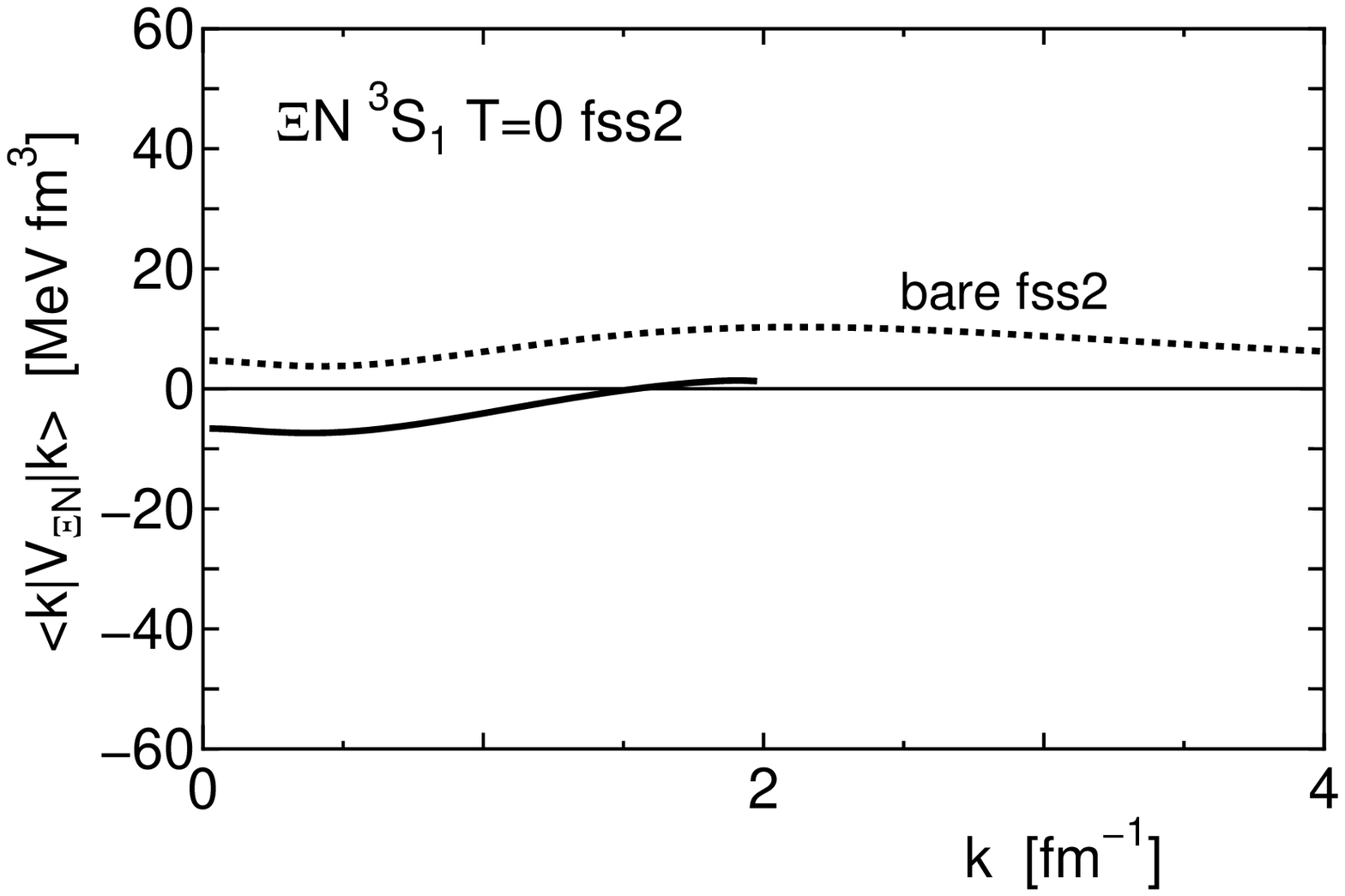}
\caption{Same as Fig. 9, but for the $\Xi N$ $^3S_1$ $T=0$ partial wave.
In this partial wave, there is no baryon-channel coupling.
}
\end{figure}

\begin{figure}[ht]
\epsfxsize=8cm
\epsfbox{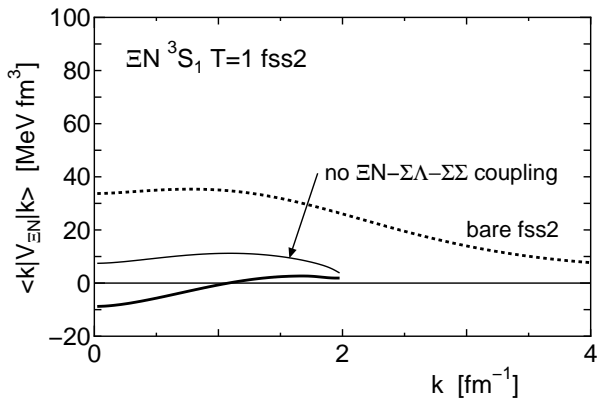}
\caption{Same as Fig. 9, but for the $\Xi N$ $^3S_1$ $T=1$ partial wave.
The thin curve shows results without taking into account the
$\Xi N$-$\Sigma\Lambda$-$\Sigma\Sigma$ channel coupling.
}
\end{figure}

There are two kinds of sources for the experimental information about the baryon-baryon
interaction in the strangeness $S=-2$ sector. One is the estimation of a $\Xi$-nucleus
s.p. potential from DWIA analyses \cite{KHA} of the $(K^-,K^+)$ $\Xi^-$ formation spectra
on $^{12}$C. The tentative conclusion is that the depth of the s.p. potential is about 14 MeV.
The other clue is the strength of the $\Lambda\Lambda$ attraction through binding
energies of double-$\Lambda$ hypernuclei.
The recent discovery of $_{\Lambda\Lambda}^{\ \ 6}$He \cite{TAK} has indicated that the
attraction of the $\Lambda\Lambda$ interaction is rather weak. There is also an
attempt \cite{AHN} to measure the $\Xi N \rightarrow \Lambda\Lambda$ inelastic
cross section and deduce the $\Lambda\Lambda$ correlation in the $\Xi$ formation
reaction. Nevertheless all these experimental data are in the rudimentary stage and far
from sufficiently constraining the interactions which have a variety of baryon-channel couplings.
Thus the investigation using theoretical models of the $\Xi N$ interaction is important.

Because the computer code for the Nijmegen $\Xi N$ interaction was not available, we present
results only with the Kyoto-Niigata quark model potential fss2 \cite{FSN}. Figure 9 shows the
low-momentum space equivalent $\Xi N$ interaction in the $^1S_0$ $T=0$ state with $\Lambda=2$
fm$^{-1}$ together with the bare matrix elements up to $k=4$ fm$^{-1}$. If the coupling to the
$\Lambda\Lambda$-$\Sigma\Sigma$ channel is neglected, the diagonal matrix element
of the equivalent interaction in the low-momentum space is close to that of the bare interaction,
which indicates that the correlation in the $\Xi N$-$\Xi N$ channel itself is rather weak. The result of
the large attraction in the $\Xi N$ interaction in the low-momentum space originates from the
baryon-channel coupling effect. Among other $S$-waves, this channel turns out to be
most attractive. As noted in the previous section, however, it is necessary to explicitly treat
the $\Xi N$-$\Lambda \Lambda$-$\Sigma\Sigma$ coupling in the $P$ space in order to obtain more
physically meaningful information.

The $^1S_0$ $T=1$ state is shown in Fig. 10(a). In this state the $\Sigma\Sigma$ channel
is not allowed and the the $\Xi N$ state couples only with the $\Sigma\Lambda$ state.
Looking at the matrix elements at large $k$, we observe that the the short-range repulsion
is weak in this $\Xi N$-$\Xi N$ channel. It is found in this channel that because
the baryon-coupling effect is strong, the sign of the diagonal matrix elements does not
indicate the property of the interaction. For example, we show, in Fig. 10(b), the diagonal
matrix elements of the low-momentum equivalent interaction with the cut-off value of
$\Lambda=3$ fm$^{-1}$. In this case $\Xi N$-$\Xi N$ diagonal matrix elements are
mostly positive. In between $\Lambda=3$ fm$^{-1}$ and 2 fm$^{-1}$, the diagonal matrix
elements change sign, which suggests that the baryon-channel coupling effects in
the $P$ space are decisively important, and thus the $\Xi N$ and $\Sigma\Lambda$
states are strongly mixed. 

The $^3S_1$ $T=0$ state is classified to the pure $(11)_a$ state in the flavor  SU$_3$
symmetry and no baryon-channel coupling appears in this state. The quark model \cite{FSN}
predicts that the bare $\Xi N$ interaction is already weak.
Figure 11 shows that the low-momentum equivalent $\Xi N$
interaction in this partial wave becomes slightly attractive.

As is shown in Fig. 12, the quark model potential fss2 predicts that the $\Xi N$ interaction
in the $^3S_1$ $T=1$
state is also not so strong. Due to the $\Xi N$-$\Sigma\Lambda$-$\Sigma\Sigma$
coupling the diagonal matrix elements of the low-momentum equivalent $\Xi N$ interaction
can be slightly negative.

Altogether, the quark model potential fss2 \cite{FSN} predicts that the $\Xi N$ interactions
in $^3S_1$ channels are weak. For the estimation of the $\Xi$-nucleus s.p. potential in
nuclear medium, we expect an attractive contribution from the $^1S_0$ $T=0$ state but a repulsive
contribution from the $^1S_0$ $T=1$ state. Higher partial waves may play an important
role for the $\Xi$-nucleus s.p. potential.

\section{Conclusions}
We have calculated low-momentum space equivalent interactions for hyperon-nucleon
interactions, starting from two models of the bare potentials, the quark model
potential fss2 \cite{FSN} and the Nijmegen OBEP model NSC97f \cite{NSC}.
An effective interaction in a restricted space which
reproduces the same eigenvalues or $T$-matrices in that space as those of the
original full-space interaction is named as an equivalent interaction in this paper.

The quark model potential uses a RGM framework, thus the
naive definition of the baryon-baryon interaction on the basis of the RGM Born kernel
is energy-dependent. Recently, the technique to eliminate the energy-dependence
was developed \cite{SUZ}. We have used this prescription for the quark model potentials
for the octet baryon-baryon interactions. The calculation of the low-momentum space
equivalent interaction for the quark model potential is interesting in two aspects.
One is related to the different non-local character from that of the one-boson exchange
picture, especially at short distance due to the RGM treatment of two quark-composite
clusters. Another point is
that the extension of the potential to the strangeness $S=-1$ and $S=-2$ sectors
based on the parameters fixed in the $NN$ sector has been shown \cite{FSN}
to be less ambiguous than the OBEP model. It is useful to elucidate the
similarity and the difference in the $\Lambda N$, $\Sigma N$ and $\Xi N$ interactions
between the quark model and the Nijmegen model. As the representative potential for
the latter model, we have employed the NSC97f \cite{NSC}, which was used in the similar
calculations recently reported by Schaefer {\it et al.} \cite{SWW}.

The merit of considering the low-momentum space equivalent interaction for the
$YN$ interactions is to eliminate the model dependent characters
at short distance, as has been demonstrated for the $NN$ case by Bogner {\it et al.}
\cite{BOG}. Thus we can concentrate on features of the $YN$ interaction
relevant to low-energy experimental hypernuclear observables.

First we have presented the $NN$ results for the sake of the comparison. The quark model
potential gives almost the same results as other modern realistic $NN$ potentials, as it should.
The $\Lambda N$ equivalent interaction in the low-momentum space turns out to be
almost identical in the quark model and the OBEP model.
This is probably because we have some amount of constraints from $\Lambda$
hypernuclear data in addition to old scattering data.

Experimental data for the $\Sigma N$ interaction is very limited. Thus the model
dependence in the theoretical construction of the $\Sigma N$ potential is large.
In fact, the quark model and the Nijmegen OBEP model
provide different equivalent interactions in the low-momentum space. In particular,
we note that the the quark model predicts repulsion in the $^3S_1$ $T=3/2$
state which originates from the Pauli effect on the quark level,
while the NSC97f potential expects attraction. On the other hand the repulsive
contribution of NSC97f in the $^1S_0$ $T=1/2$ state is rather strong.

The $\Xi N$ interaction is less known experimentally. This subject is one of the
primary subjects of the on-going J-PARC project \cite{JPC}. The matrix elements
in the low-momentum space from the quark model potential suggests that
the $\Xi$-nucleus s.p. potential is probably attractive due to the interaction
in the $^1S_0$ state in the isospin $T=0$ channel, but weak.
It is expected that there are strong baryon-channel coupling effects both in the
high-momentum space and in the low-momentum space. In this paper, we do not
consider higher partial waves. Actually they may be important for the
quantitative discussion of the $\Xi$ hyperon in nuclear medium. These are the future
subjects to study.

In hyperon-nucleon interactions, baryon-channel couplings appear in most cases.
If the coupling effect is important, as in the case of the $\Sigma N$ $^3S_1$ $T=1/2$
state in the Nijmegen NSC97f model and the $\Xi N$ $^1S_0$ $T=1$ state in the
quark model, we have to solve the channel-coupling problem in the low-momentum
space in order to obtain physically meaningful quantities. When considering
hyperons in nuclear medium, we also have to take into account the Pauli exclusion
effect on the hadron level in a nuclear many body system. The effect of the three-body
correlation through the coupling of the Pauli effect and the elimination of the high
momentum components of the interaction may be important for hyperons in nuclear medium.
\bigskip

This study is supported by Grants-in-Aid for Scientific
Research (C) from the Japan Society for the Promotion of
Science (Grant Nos. 17540263 and 18540261).


\begin{thebibliography}{99}
\bibitem{BOG} S.K. Bogner, T.T.S. Kuo, and A. Schwenk, Phys. Rep. {\bf 386}, 1 (2003).
\bibitem{EP99} E. Epelbaum, W. Gl\"{o}ckle, A. Kr\"{u}ger, and Ulf-G. Mei{\ss}ner,
Nucl. Phys. {\bf A645}, 413 (1999).
\bibitem{OKU} S. Okubo, Prog. Theor. Phys. {\bf 12}, 603 (1954)
\bibitem{BH58} C. Bloch and J. Horowitz, Nucl. Phys. {\bf 8}, 91 (1958).
\bibitem{FB58} H. Feshbach, Ann. Phys. {\bf 5}, 357 (1958).
\bibitem{BRAN} B.H. Brandow, Rev. Mod. Phys. {\bf 39}, 771 (1967).
\bibitem{SL80} K. Suzuki and S.Y. Lee, Prog. Theor. Phys. {\bf 64}, 2091 (1980).
\bibitem{LS80} S.Y. Lee and K. Suzuki, Phys. Lett. {\bf 91B}, 173 (1980).
\bibitem{NIJDF} M.M. Nagels, T.A. Rijken and J.J. de Swart,
Phys. Rev. D {\bf 12}, 744(1975);
 D {\bf 15} (1977), 2547; D {\bf 20}, 1633 (1979).
\bibitem{NIJNS} P.M.M. Maessen, T.A. Rijken and J.J. de Swart,
Phys. Rev. C {\bf 40}, 2226 (1989).
\bibitem{NSC} T.A. Rijken, V.G.J. Stoks, and Y. Yamamoto, Phys. Rev. {\bf C59}, 21 (1999).
\bibitem{HHS} B. Holzenkamp, K. Holinde and J. Speth, Nucl. Phys. {\bf A500}, 485 (1989).
\bibitem{RHS} A. Reuber, K. Holinde and J. Speth, Nucl. Phys. {\bf A570}, 543 (1994).
\bibitem{HM} J. Haidenbauer and Ulf-G. Mei{\ss}ner, Phys. Rev. {\bf C72}, 044005 (2005).
\bibitem{FU96a} Y. Fujiwara, C. Nakamoto and Y. Suzuki, Phys. Rev. Lett.
{\bf 76}, 2242 (1996).
\bibitem{FU96b} Y. Fujiwara, C. Nakamoto and Y. Suzuki,
Phys. Rev. C {\bf 54}, 2180 (1996).
\bibitem{FSN} Y. Fujiwara, Y. Suzuki, and.C. Nakamoto, Prog. Part. Nucl. Phys. {\bf 58}, 439
(2007).
\bibitem{CEFT1} H. Polinder, J. Haidenbauer and Ulf-G. Mei{\ss}ner, Nucl. Phys. {\bf A779}, 244 (2006).
\bibitem{CEFT2} H. Polinder, J. Haidenbauer and Ulf-G. Mei{\ss}ner, nucl-th/0705.3753.
\bibitem{SWW} B.-J. Schaefer, M. Wagner, J. Wambach, T.T.S. Kuo, and G.E. Brown,
Phys. Rev. {\bf C73}, 011001(R) (2006).
\bibitem{SUZ} Y. Suzuki, H. Matsumura, M. Orabi,Y. Fujiwara, P. Descouvemont,
M. Theeten and D. Baye, in preparation.
\bibitem{FKNS} Y. Fujiwara, M. Kohno, C. Nakamoto and Y. Suzuki, Prog. Theor. Phys. {\bf 104},
1025 (2000).
\bibitem{SZ82} K. Suzuki, Prog. Theor. Phys. {\bf 68}, 246 (1982).
\bibitem{SO83} K. Suzuki and R. Okamoto, Prog. Theor. Phys. {\bf 70}, 439 (1983).
\bibitem{SOEK} K. Suzuki, R. Okamoto, P. J. Ellis, T. T. S. Kuo, Nucl. Phys. {\bf A567}, 576 (1994).
\bibitem{FUJI} S. Fujii, E. Epelbaum, H. Kamada, R. Okamoto, K. Suzuki and W. Gl\"{o}ckle,
Phys. Rev. {\bf C70}, 024003 (2004).
\bibitem{OFS} R. Okamoto, S. Fujii and K. Suzuki, Int. J. Mod. Phys. {\bf 14}, 21 (2005).
\bibitem{TAM} O. Hashimoto and H. Tamura, Prog. Part. Nucl. Phys. {\bf 57}, 564 (2006).
\bibitem{OSY} M. Oka, K. Shimizu and K. Yazaki, Nucl. Phys. {\bf A464}, 700 (1987).
\bibitem{HAR} T. Harada, S. Shinmura, Y. Akaishi, and H. Tanaka, Nucl. Phys.
{\bf A507}, 715 (1990).
\bibitem{HAY} R.S. Hayano {\it et al.}, Phys. Lett. {\bf B231}, 355 (1989).
\bibitem{NAG} T. Nagae {\it et al.}, Phys. Rev. Lett. {\bf 80}, 1605 (1998).
\bibitem{BNL} S. Bart {\it et al.}, Phys. Rev. Lett. {\bf 83}, 5238 (1999).
\bibitem{NOUM} H. Noumi {\it et al.}, Phys. Rev. Lett. {\bf 89}, 072301 (2002); {\bf 90},
049902(E) (2003).
\bibitem{HH} T. Harada and Y. Hirabayashi, Nucl. Phys. {\bf A759}, 143 (2005).
\bibitem{MK} M. Kohno, Y. Fujiwara, Y. Watanabe, K. Ogata and M. Kawai,
Phys. Rev. {\bf C74}, 064613 (2006).
\bibitem{KF} M. Kohno, Y. Fujiwara, T. Fujita, C. Nakamoto and Y. Suzuki, Nucl. Rev.
{\bf A674}, 229 (2000).
\bibitem{KHA} P. Khaustov  {\it et al.}, Phys. Rev. {\bf C61}, 054603 (2000).
\bibitem{TAK} H. Takahashi {\it et al.}, Phys. Rev. Lett. {\bf 87}, 212502 (2001).
\bibitem{AHN} J.K. Ahn {\it et al.}, Phys. Lett. {\bf B633}, 214 (2006).
\bibitem{JPC} T. Nagae, {\it Proc. of Int. Nucl. Phys. Conf.}, Tokyo, Japan, 3-8 June, 2007.
\end{thebibliography}
\end{document}